# Silicon Sovereigns: Artificial Intelligence, International Law, and the Tech-Industrial Complex

Simon Chesterman*

*Review Essay*

*AI Needs You*. By Verity Harding. Princeton: Princeton University Press, 2024. Pp. x, 274. Index.

*Technological Internationalism and World Order*. By Waqar H. Zaidi. Cambridge: Cambridge University Press, 2023. Pp. xiv, 300. Index.

*Power and Progress: Our 1000-Year Struggle Over Technology and Prosperity*. By Daron Acemoglu & Simon Johnson. New York: Public Affairs, 2023. Pp. xv, 546. Index.

Debates over the governance of artificial intelligence (AI) tend to assume that it will be important and transformative across many areas of human endeavor.[1] Though filtering out the hype can be challenging,[2] the last round of Nobel Prizes supports at least some of these claims: the prize in physics was awarded for foundational work on machine learning, while the one for chemistry was awarded to researchers using AI to model protein folding.[3] The focus in such debates is typically over how to reap the benefits of AI while minimizing or mitigating known and unknown risks.[4] Yet, the question of how those benefits and risks will be *distributed* — who will win and who will lose — is less commonly articulated.

---

* Vice Provost and David Marshall Professor of Law, National University of Singapore. Many thanks to Fakhar Abbas, Viv Chesterman, Jungpil Hahn, Shaleen Khanal, Tristan Koh, Ernest Lim, Valerie Lim, Renae Loh, Hakim Norhashim, Eric Orlowski, Eka Nugraha Putra, Shrestha Saha, Araz Taeihagh, Ming Tan, Tan Hsien-Li, Jun Yu, and Audrey Yue for comments on an earlier draft. Errors, omissions, and hallucinations are attributable to the author alone.

[1] The term "artificial intelligence" is used in this essay to refer to systems that can apply cognitive functions to specific tasks typically undertaken by a human. For a discussion of attempts to define AI, see STUART J. RUSSELL AND PETER NORVIG, ARTIFICIAL INTELLIGENCE: A MODERN APPROACH 1-5 (3rd ed. 2010). Four broad definitional approaches can be identified: acting humanly (the famous Turing Test), thinking humanly (modelling cognitive behavior), thinking rationally (building on the logicist tradition), and acting rationally (a rational-agent approach favored by Russell and Norvig as it is not dependent on a specific understanding of human cognition or an exhaustive model of what constitutes rational thought).

Techno-utopians enthuse that everyone will win: the pie will be bigger; the rising tide will lift all boats.[5] Concerns about inequality or the environmental impact of AI[6] are batted aside with the promise that AI itself will solve such problems.[7] Others, including a surprising fraction of those developing AI systems themselves, warn of darker, dystopian futures in which AI turns on humanity, either through misalignment of objectives or the emergence of a superintelligence that regards its creators in the way that we might regard lesser creatures such as dogs — or ants.[8] Everyone loses.

---

[2] ARVIND NARAYANAN AND SAYASH KAPOOR, AI SNAKE OIL: WHAT ARTIFICIAL INTELLIGENCE CAN DO, WHAT IT CAN'T, AND HOW TO TELL THE DIFFERENCE (2024).

[3] Nobel Prizes 2024, *at* https://www.nobelprize.org/all-nobel-prizes-2024. Goldman Sachs has estimated that by 2034 U.S. GDP will increase by more than 2 percent as a result of AI. AI May Start to Boost U.S. GDP in 2027 (Goldman Sachs Research, 7 November 2023), *at* https://www.goldmansachs.com/insights/articles/ai-may-start-to-boost-us-gdp-in-2027. McKinsey's Global Institute goes further and predicts a 5 to 13 percent boost by 2040. Michael Chui et al., The Economic Potential of Generative AI: The Next Productivity Frontier (McKinsey, San Francisco, 14 June 2023), *at* https://www.mckinsey.com/capabilities/mckinsey-digital/our-insights/the-economic-potential-of-generative-ai-the-next-productivity-frontier.

[4] *See, e.g.,* RYAN ABBOTT, THE REASONABLE ROBOT: ARTIFICIAL INTELLIGENCE AND THE LAW (2020); MICHAEL GUIHOT AND LYRIA BENNETT MOSES, ARTIFICIAL INTELLIGENCE, ROBOTS AND THE LAW (2020); MATTHEW LAVY AND MATT HERVEY, THE LAW OF ARTIFICIAL INTELLIGENCE (2020); FRANK PASQUALE, NEW LAWS OF ROBOTICS: DEFENDING HUMAN EXPERTISE IN THE AGE OF AI (2020); SIMON CHESTERMAN, WE, THE ROBOTS? REGULATING ARTIFICIAL INTELLIGENCE AND THE LIMITS OF THE LAW (2021); DOMINIKA EWA HARASIMIUK AND TOMASZ BRAUN, REGULATING ARTIFICIAL INTELLIGENCE: BINARY ETHICS AND THE LAW (2021); JINGHAN ZENG, ARTIFICIAL INTELLIGENCE WITH CHINESE CHARACTERISTICS: NATIONAL STRATEGY, SECURITY AND AUTHORITARIAN GOVERNANCE (2022); MARK CHINEN, THE INTERNATIONAL GOVERNANCE OF ARTIFICIAL INTELLIGENCE (2023); THE CAMBRIDGE HANDBOOK ON LAW, REGULATIONS, AND POLICY FOR HUMAN-ROBOT INTERACTION (Woodrow Barfield, Ugo Pagallo, and Yueh-Hsuan Weng eds., 2024); ANDEED MA, JAMES ONG, AND SIOK SIOK TAN, AI FOR HUMANITY: BUILDING A SUSTAINABLE AI FOR THE FUTURE (2024); TSHILIDZI MARWALA, THE BALANCING PROBLEM IN THE GOVERNANCE OF ARTIFICIAL INTELLIGENCE (2024); International AI Safety Report 2025 (Department for Science, Innovation and Technology and AI Safety Institute, London, 29 January 2025), *at* https://www.gov.uk/government/publications/international-ai-safety-report-2025.

[5] *See, e.g.,* RAY KURZWEIL, THE SINGULARITY IS NEAR: WHEN HUMANS TRANSCEND BIOLOGY (2005); YUVAL NOAH HARARI, HOMO DEUS: A BRIEF HISTORY OF TOMORROW (2016); REID HOFFMAN AND GREG BEATO, SUPERAGENCY: WHAT COULD POSSIBLY GO RIGHT WITH OUR AI FUTURE (2025).

[6] *See, e.g.,* KATE CRAWFORD, THE ATLAS OF AI: POWER, POLITICS, AND THE PLANETARY COSTS OF ARTIFICIAL INTELLIGENCE (2021).

[7] *Former Google CEO Eric Schmidt Urges AI Acceleration, Dismisses Climate Goals*, COMPUTING UK, 8 October 2024; ; Dario Amodei, Machines of Loving Grace: How AI Could Transform the World for the Better (Anthropic, San Francisco, October 2024), *at* https://darioamodei.com/machines-of-loving-grace.

Between the extremes are those trying to think through where the gains and losses of AI will fall. In realist circles, it has become common to speak of AI in the language of an arms race, a comfortingly familiar frame that pits the West against a rising China.[9] Anu Bradford's *Digital Empires*, for example, posits a battle for regulatory pre-eminence between the market-based U.S. model, a rights-based approach favored in the European Union, and China's state-centric regime. At stake, she argues, is nothing less than the "soul of the digital economy."[10] Her book captures the geopolitical moment, in particular the Sino-U.S tensions playing out over access to high-performance computing power.[11] Arch-realist Henry Kissinger — unable to step away from the game of *realpolitik* even as he approached a hundred years of age — spent his final months writing on AI and warning of its geopolitical implications.[12]

An alternative framing adopts a North-South axis, noting the 750 million people without stable electricity and the more than two billion unconnected to the internet.[13] A report by the United Nations AI Advisory Body examined some of the most prominent efforts at international governance of AI and found that seven countries (the wealthy, industrialized members of the G7) are party to all of them, while more than a hundred other states are party to none.[14] For all the worry about misuse of AI, many in developing countries are more concerned about *missed* uses and being left behind.

---

[8] Nick Bostrom, Superintelligence: Paths, Dangers, Strategies (2014); Robert Sparrow, *Friendly AI Will Still Be Our Master*, 39(5) AI & Society 2439 (2023); Joud Mohammed Alkhalifah et al., *Existential Anxiety About Artificial Intelligence (AI): Is It the End of Humanity Era or a New Chapter in the Human Revolution*, 15 Frontiers in Psychiatry 1368122 (2024).

[9] Kai-Fu Lee, AI Superpowers: China, Silicon Valley, and the New World Order (2018); Andrew R. Chow and Billy Perrigo, *The AI Arms Race Is Changing Everything*, Time, 16 February 2023. *Cf.* Kerry McInerney, *Yellow Techno-Peril: The 'Clash of Civilizations' and Anti-Chinese Racial Rhetoric in the US–China AI Arms Race*, 11(2) Big Data & Soc'y (2024).

[10] Anu Bradford, Digital Empires: The Global Battle to Regulate Technology 385 (2023).

[11] *Cf.* Chris Miller, Chip War: The Fight for the World's Most Critical Technology (2022); *The United States Announces Export Controls to Restrict China's Ability to Purchase and Manufacture High-End Chips*, 117(1) Am. J. Int'l L. 144 (2023).

[12] Henry Kissinger et al., The Age of AI: And Our Human Future (2021); Henry Kissinger, Eric Schmidt, and Craig Mundie, Genesis: Artificial Intelligence, Hope, and the Human Spirit (2024).

[13] Population of Global Offline Continues Steady Decline to 2.6 Billion People in 2023 (International Telecommunications Union, Geneva, 12 September 2023), *at* https://www.itu.int/en/mediacentre/Pages/PR-2023-09-12-universal-and-meaningful-connectivity-by-2030.aspx.

Yet, the most important divide may not be East-West or North-South but public-private. For AI is shifting economic and, increasingly, political power away from governments.[15] That is most obvious in the deployment of ever more powerful products with minimal regulatory oversight. But it is also true at the level of fundamental research. Machine learning models that power systems like ChatGPT began in publicly-funded universities; by 2022, of the dozens of significant models tracked by Stanford's AI index, all but three were released by industry.[16] Outside of the European Union, states have been wary of introducing new laws to regulate AI for fear of losing a competitive advantage or driving innovation elsewhere.[17] Efforts to regulate technology companies by applying existing laws — prominently including antitrust and intellectual property — have seen government lawyers hopelessly outgunned by their corporate counterparts.[18]

All of this poses a challenge to public lawyers generally and international lawyers in particular. In structural terms, if the twentieth century saw a turn from bilateralism to multilateralism, with the emergence of truly international organizations, the twenty-first century may be witnessing a fracturing of those structures — and indeed a decline in the preeminence of states as the primary political vehicle on the global stage. Though corporatist critiques of the modern political order are hardly new,[19] the nature and scale of the power wielded by today's tech giants rivals the role occupied by the East India Company in the early nineteenth century, when it controlled half of global trade and had its own army.[20] Today's tech behemoths may lack that measure of economic or military power, but their global cultural and political influence is arguably greater.[21]

---

[14] UN AI Advisory Body, Governing AI for Humanity: Final Report (United Nations, New York, September 2024), *at* https://www.un.org/en/ai-advisory-body, at 9. Disclosure: the author served as Principal Researcher for the UN AI Advisory Body at the time the report was drafted. *Cf.* Heidi Aly, *Digital Transformation, Development and Productivity in Developing Countries: Is Artificial Intelligence a Curse or a Blessing?*, 7(4) Rev. Eco. & Pol. Sci. (REPS) 238 (2022)

[15] A counterargument can be made that some authoritarian governments use AI and/or co-opt technology companies to reinforce their power, though this typically takes the form of outsourcing government functions or relying on third-party tools to engage in surveillance or the suppression of dissent. *See, e.g.,* Steven Feldstein, The Rise of Digital Repression: How Technology Is Reshaping Power, Politics, and Resistance (2021).

[16] AI Index Report 2023 (Institute for Human-Centered AI, Stanford University, Stanford, CA, April 2023), *at* aiindex.stanford.edu/wp-content/uploads/2023/04/HAI_AI-Index-Report_2023.pdf, at 50. A proxy for the computing power at an organization's disposal is the number of parameters. Industry models in 2021 were on average 29 times larger than those in academic institutions. Nur Ahmed, Muntasir Wahed, and Neil C. Thompson, *The Growing Influence of Industry in AI Research*, 379(6635) Science 884 (2023). Private investment in AI in 2022 was eighteen times greater than in 2013. In 2021, the U.S. government allocated US$1.5 billion to non-defense academic research into AI; Google spent that much on DeepMind alone. Talent has followed. The number of AI research faculty in universities has not risen significantly since 2006, while industry positions have grown eightfold. Two decades ago, only about twenty percent of graduates with a PhD in AI went to industry; today around seventy percent do.

[17] Simon Chesterman, *From Ethics to Law: Why, When, and How to Regulate AI*, *in* The Handbook on the Ethics of AI 113 (David J. Gunkel ed., 2024).

[18] *See* Shaleen Khanal, Hongzhou Zhang, and Araz Taeihagh, *Why and How Is the Power of Big Tech Increasing in the Policy Process? The Case of Generative AI*, forthcoming Pol. & Soc. (2025).

Another dimension of the governance challenge posed by AI is time. The speed with which innovations and new capacities are now launched recalls the Red Queen's advice to Alice in Wonderland: just to stay in one place you have to run as fast as you can; if you want to get somewhere else, you must run at least twice as fast as that.[22] Keeping up with the pace of change is reflected in the academic work of computer scientists, whose stock in trade is not books or even journal articles, but conference proceedings.[23] A relentless presentism can lead observers to overestimate the impact of a new technology in the short-term even as they underestimate it in the long-term, a phenomenon sometimes termed "Amara's law."[24]

---

[19] *See, e.g.,* SUSAN STRANGE, THE RETREAT OF THE STATE: THE DIFFUSION OF POWER IN THE WORLD ECONOMY (1996); STEPHEN D. KRASNER, SOVEREIGNTY: ORGANIZED HYPOCRISY (2001); CHRISTOPHER N. MAY, GLOBAL CORPORATIONS IN GLOBAL GOVERNANCE (2015).

[20] *See* ANTONY ANGHIE, IMPERIALISM, SOVEREIGNTY, AND THE MAKING OF INTERNATIONAL LAW (2005); H.V. BOWEN, THE BUSINESS OF EMPIRE: THE EAST INDIA COMPANY AND IMPERIAL BRITAIN, 1756-1833 (2006); EMILY ERIKSON, BETWEEN MONOPOLY AND FREE TRADE: THE ENGLISH EAST INDIA COMPANY, 1600-1757 (2017).

[21] Elon Musk, to pick an obvious example, has extraordinary leverage through his ownership of SpaceX, which dominates space exploration and satellite internet, Tesla, which revolutionized the electric vehicle market, and X (née Twitter), which offers a global platform for his views. Ronan Farrow, *Elon Musk's Shadow Rule*, NEW YORKER, 21 August 2023; WALTER ISAACSON, ELON MUSK (2023).

[22] LEWIS CARROLL, THROUGH THE LOOKING-GLASS ch 2 (1872).

[23] Jinseok Kim, *Author-Based Analysis of Conference Versus Journal Publication in Computer Science*, 70(1) J. ASS'N INFO. SCI. & TECH. 71 (2019).

[24] The late futurist Roy Amara is typically credited with coining the term. *See, e.g.,* DOC SEARLS, THE INTENTION ECONOMY: WHEN CUSTOMERS TAKE CHARGE 257 (2012). *Cf.* J.C.R. LICKLIDER, LIBRARIES OF THE FUTURE 17 (1965) (referring to a "modern maxim" that states that people "tend to overestimate what can be done in one year and to underestimate what can be done in five or ten years"). For their part, governments face a particular challenge, being forced to choose between under-regulating an emerging field, possibly exposing their citizens to risk, or over-regulating, and perhaps limiting innovation or driving it elsewhere. David Collingridge described this dilemma half a century ago: In the early stages of innovation, exercising control would be easy — but not enough is known about the potential harms to warrant slowing development. By the time those harms are

Such presentism and the efforts to forecast possible futures can blind us to the realization that it may be more profitable not to look forward but to look back. This is not the first time humanity has confronted a technology with the potential for good or ill, giving rise to a clash between public interest and private ones — nor will it be the last. In distinct ways, each of the three books considered in this review essay encourages a historicist turn, situating AI and related technologies in their historical moment and seeking lessons from past technological revolutions that similarly challenged norms, even as they also revealed or exacerbated inequalities through the distribution of benefits and risks.

Verity Harding's *AI Needs You* draws on three late-twentieth-century examples: the space race, in vitro fertilization, and the internet.[25] Waqar Zaidi looks back further in that century to the emergence of aviation and atomic energy, along with the rise of what he terms "technological internationalism."[26] Daron Acemoglu and Simon Johnson propose a more ambitious sweep of a thousand years of "power and progress" in the title of their own volume, though they mostly focus on the industrial revolution onwards.[27]

None of these books explicitly foregrounds international law or institutions as the mechanism that should play a lead role in regulating emerging technologies like AI. Nevertheless, each has much to say about the possibilities and limitations of global efforts to govern them — if only through the frustration the authors variously express at the inadequacy of market- and state-based efforts. Indeed, if there is a throughline that resonates with each work, it is the catastrophic mismatch between leverage and interest as between those who could govern AI, exacerbated by the incentives that encourage technology companies to "move fast and break things"[28] while governments are left flatfooted or left behind.

---

apparent, control has become costly and difficult. David Collingridge, The Social Control of Technology 19 (1980).

[25] Verity Harding, AI Needs You: How We Can Change AI's Future and Save Our Own (2024).

[26] Waqar H. Zaidi, Technological Internationalism and World Order: Aviation, Atomic Energy, and the Search for International Peace, 1920–1950 (2023).

[27] Daron Acemoglu and Simon Johnson, Power and Progress: Our 1000-Year Struggle Over Technology and Prosperity (2023).

[28] "Move fast and break things" was an early motto at Facebook intended to push developers to take risks; the phrase appeared on office posters and featured in a letter from Mark Zuckerberg to investors when the company went public in 2012. Form S-1 Registration Statement of Facebook, Inc. (United States Securities and Exchange Commission, Washington, DC, 1 February 2012), *at* https://www.sec.gov/Archives/edgar/data/1326801/000119312512034517/d287954ds1.htm#toc287954_10. Over time, it came to be embraced as a mantra applicable to technological disruption more generally, adopted

# 1 Artificial Intelligence Needs Who?

Harding's résumé gives her a unique perspective on the various camps involved in technology policy, including stints as Global Head of Policy at Google DeepMind and as an adviser to Nick Clegg, when he was Deputy Prime Minister of Britain. She now directs the AI & Geopolitics Project at Cambridge University's Bennett Institute for Public Policy. Her stated objective is to seek lessons from past technological transformations, though she sets a somewhat arbitrary cut-off of the Second World War in terms of time and largely limits herself to the United States and Britain. Indeed, the "you" of the title is often explicitly revealed to be directed at those in Western democracies.[29]

Nonetheless, her book is an urgent challenge to the passivity with which many people and governments appear to view the development of AI today. Like it or not, we are already shaping this emergent technology. Well-known problems, such as the potential for discrimination, are not a reflection of any inherent bias on the part of our silicon creations. On the contrary, such bias is typically a faithful reflection of the data that we have, directly or indirectly, fed into these systems. AI "is not human," Harding observes. "But it is *us*."[30]

Harding chooses three technological inflection points to argue that democratic societies can enable "a myriad of citizens" to take an active role in shaping the future of AI.[31] The first is the Space Race — the Cold War rivalry between the United States and the Soviet Union as each sought dominance in orbit and beyond. Technological advances in spaceflight had clear implications for security, including the ability to launch missiles and position satellites for reconnaissance, but efforts to launch uncrewed and then crewed vessels beyond our atmosphere captured the wider public imagination.[32] Harding argues that U.S. Presidents

---

by countless Silicon Valley imitators. As Facebook matured, however, and as the potential harms caused by such disruption grew, the slogan fell from favour. JONATHAN TAPLIN, MOVE FAST AND BREAK THINGS: HOW FACEBOOK, GOOGLE, AND AMAZON CORNERED CULTURE AND UNDERMINED DEMOCRACY (2017); Hemant Taneja, *The Era of "Move Fast and Break Things" Is Over*, HARVARD BUSINESS REVIEW, 22 January 2019. See also Simon Chesterman, *"Move Fast and Break Things": Law, Technology, and the Problem of Speed*, 33 SING. ACAD. L.J. 5 (2021).

[29] HARDING, *supra* note 25, at 180 ("As leading AI democracies we must ask ourselves how our use of AI, the examples *we* set, will shape our societies, our governments, and the world at large.")

[30] *Id*. at 12. *See also* TAINA BUCHER, IF … THEN: ALGORITHMIC POWER AND POLITICS (2018).

[31] HARDING, *supra* note 25, at 26. She describes the book as "my love letter to the painstaking and generally unglamourous world of policy-making in a democracy." *Id*. at 224.

[32] ROGER D. LAUNIUS, REACHING FOR THE MOON: A SHORT HISTORY OF THE SPACE RACE (2019); ALBERT K. LAI, THE COLD WAR, THE SPACE RACE, AND THE LAW OF OUTER SPACE: SPACE FOR PEACE (2021).

Eisenhower, Kennedy, and Johnson showed a willingness to balance national defense and the "greater ideals of international cooperation and pacificism,"[33] culminating in the Outer Space Treaty of 1967.[34] AI, she argues — exaggerating, to be sure — "like space in the middle of the twentieth century, is a new frontier and a blank sheet for global norms."[35]

Her second analogy is in vitro fertilization (IVF). Following the birth of Louise Joy Brown in England in 1978, the first person conceived using the new technique, a biotechnology revolution was shaped by the "careful setting of boundaries and pursuit of consensus."[36] Of the examples considered in her book, Harding finds the greatest similarity between handwringing over AI today and the biotech debates of the 1970s and 1980s, in particular fears "about corporate influence, about an unknown future, about what it means to be human."[37] Yet the goal of fertility treatment — enabling otherwise infertile families to have a child — is unusually clear and the red lines beyond it more easily debated and drawn. In the period Harding considers, these included limits on human embryo research, with provision for licensing and a "fourteen-day rule" that banned experimentation on embryos more than two weeks after fertilization.[38] Britain's IVF legislation may have been progressive, but Harding over-eggs it somewhat as being "the most innovative and world-leading scientific regulation of the century."[39]

The third comparison is the early development of what became the internet, which she describes as "a story of convergence: between baby boomers who grew up believing the promise of 1960s liberalism and those who felt betrayed by it; between young modernizing progressive politicians and the business titans of the new Gilded Age; and between the newly developed 'internet community' and those tasked with regulating it."[40] She notes in particular the manner in which the Internet Corporation for Assigned Names and Numbers (ICANN), established a free and open global network through multistakeholder and multinational cooperation, epitomized by "unglamorous efforts by normal people in meeting rooms trying to make things work."[41]

---

[33] HARDING, *supra* note 25, at 26.

[34] Treaty on Principles Governing the Activities of States in the Exploration and Use of Outer Space, including the Moon and Other Celestial Bodies (Outer Space Treaty), done at London, Moscow, and Washington, 27 January 1967 (in force 10 October 1967), *at* http://www.oosa.unvienna.org. *See also* Tanja Masson-Zwaan and Roberto Cassar, *The Peaceful Uses of Outer Space*, *in* THE OXFORD HANDBOOK OF UNITED NATIONS TREATIES 181 (Simon Chesterman, David M. Malone, and Santiago Villalpando eds., 2019).

[35] HARDING, *supra* note 25, at 65. Neither outer space nor AI were ever a truly "blank sheet." *See, e.g.,* Vladlen S. Vereshchetin and Gennady M. Danilenko, *Custom as a Source of International Law of Outer Space*, 13(1) J. SPACE L. 22 (1985).

[36] HARDING, *supra* note 25, at 26.

[37] *Id*. at 76.

[38] *Id*. at 95.

[39] *Id*. at 92.

These are curious examples on which to rest an argument for wider and more participatory public involvement in charting the path of new technologies. Space exploration is today driven precisely by private interests, with the U.S. space program dominated by a single individual — Elon Musk — who also happens to be a leading figure in AI.[42] Reproductive rights have become some of the most politically divisive issues in the United States.[43] As for governance of the internet, decentralized control might preserve freedom and openness while encouraging innovation, but that openness has also allowed the proliferation of tools that enhance surveillance, monetize human attention, and replace human labor.[44]

Nonetheless, Harding's central message is that the goal of developing AI that recognizes our weaknesses, aligns with our strengths, and serves the public good requires greater participation. Her book is most compelling in its argument that the future cannot be left to the innovators and disruptors alone. Unfortunately, she concedes (drawing on personal experience), compromise, humility, and "acceptance that your world view might not be correct" are qualities not found in abundance in the tech industry.[45]

---

[40] *Id*. at 126.

[41] *Id*. at 27.

[42] Lewis D. Solomon, The Privatization of Space Exploration: Business, Technology, Law and Policy (2008); Chad Anderson, *Rethinking Public–Private Space Travel*, 29(4) Space Policy 266 (2013); Johan Eriksson and Lindy M. Newlove-Eriksson, *Outsourcing the American Space Dream: SpaceX and the Race to the Stars*, 21(1) Astropolitics 46 (2023).

[43] Mary Ziegler, Reproduction and the Constitution in the United States (2022); Elyshia Aseltine and Sheldon Ekland Olson, Abortion in the United States: The Moral and Legal Landscape (2024).

[44] *See, e.g.,* Araz Taeihagh, *Governance of Artificial Intelligence*, 40 Pol. & Soc. 137 (2021); Markus Furendal and Karim Jebari, *The Future of Work: Augmentation or Stunting?*, 36(36) Phil. & Tech. (2023); Jochen Wirtz et al., *Corporate Digital Responsibility in Service Firms and Their Ecosystems*, 26(2) Journal of Service Research 173 (2023). The innovative multi-stakeholder entity ICANN may embody transparent deliberation and decision-making, but it is with regard to an exceptionally narrow and easily defined category of disputes over unique identifiers on the internet.

[45] Harding, *supra* note 25, at 120.

The absence of those qualities amplified the breakdown of trust that she documents, starting with Snowden revelations around 2013 that showed even democracies were abusing the surveillance potential of the digital world. That, in turn, was compounded by the realization that the economic model of this new world relies on corporations harvesting vast amounts of data also: "Artificial Intelligence is being built in an environment of shattered trust — between citizens and their governments, between governments and powerful corporations, and between political ideologies of liberal democracy and authoritarianism."[46]

Despite the fact that the title of her book and its closing words are a call to action to the global (or at least Western) "you," there is also a strong thread of the "great man of history" to Harding's account. In her telling, the success of the Space Race depended on exceptional leadership by three American presidents. IVF's threading of a needle of compromise was due in significant part to personal qualities of Baroness Warnock and Prime Minister Margaret Thatcher. And much of the discussion of the early internet focuses on the pivotal role played by the junior senator from Tennessee, Al Gore.

Indeed, one of Harding's first lessons is the importance of "powerful political leadership to exert influence over the future direction of technology, and humanity."[47] Today, "while we have the technological power to lead, there is neither the political will nor capacity to do so in a way that could generate benefits for humanity worldwide."[48]

Unfortunately, Donald Trump's second presidential term offers little hope of addressing such an abdication of leadership — on the contrary, he appears to have doubled down on it with his elevation of tech titans above his own cabinet at his inauguration and offering Elon Musk office space adjacent to the White House.[49] One of Trump's first acts as President was to roll back even the thin gruel of his predecessor's executive order that had sought to bypass

---

[46] *Id*. at 210–11. The Cambridge Analytica scandal of 2016 and the possibility that it affected that year's U.S. presidential election started a flurry of efforts in the area of AI governance. Simon Chesterman et al., *The Evolution of AI Governance*, 57(9) IEEE COMPUTER 80 (2024).

[47] HARDING, *supra* note 25, at 65.

[48] *Id*. at 38.

[49] Theodore Schleifer and Madeleine Ngo, *Elon Musk and His Allies Storm Into Washington and Race to Reshape It*, N.Y. TIMES, 29 January 2025.

legislative deadlock and address at least a few of the risks posed by AI through administrative action.[50]

In a section towards the end of her book, aptly titled "The Red, White, and Blue Elephant in the Room,"[51] Harding argues that, half a century after the Outer Space Treaty was signed, the United States has another chance to lead. She invokes the advice Warren Buffett famously offered the musician Bono, who was seeking a strategy to convince the United States to support funding to fight AIDS in Africa: "Don't appeal to the conscience of America, appeal to its greatness."[52] For the time being, however, with regard to technology policy at least, the United States appears content to follow.

# 2  The Rise and Fall of Technological Internationalism

The isolationism of Donald Trump is hardly unique in U.S. politics. George Washington himself famously abjured foreign entanglements in his farewell address, though subsequent presidents quickly discovered that was easier said than done.[53] Indeed, his successors were among the great architects of the modern international order — Woodrow Wilson and the League of Nations, Franklin Roosevelt and the UN.[54] Waqar Zaidi taps into this theme in his chronicle of efforts to internationalize two technologies that came to dominate twentieth century warfare: aviation and atomic energy.[55] Each case saw efforts, notably on the part of U.S. and British internationalists, to take them out of government hands and put them under the control of international organizations. Based on a doctoral thesis completed at Imperial College London under the supervision of historian David Edgerton, Zaidi keeps his focus narrow. Indeed, the term "artificial intelligence" appears but once on the very last page of the book — a single sentence drily noting that, like aviation and atomic energy, AI today offers economic growth, arms races, "and possibly the extinction of the human race."[56]

---

[50] Executive Order on Safe, Secure, and Trustworthy Development and Use of Artificial Intelligence (Executive Order 14110) 2023 (U.S.); Cary Coglianese, *People and Processes: AI Governance Under Executive Order 14110*, 49(1) ADMINISTRATIVE & REGULATORY LAW NEWS 9 (2023).

[51] HARDING, *supra* note 25, at 231.

[52] Madeleine Bunting, *Bono Talks of US Crusade*, GUARDIAN, 16 June 2005.

[53] JACK GODWIN, THE ARROW AND THE OLIVE BRANCH: PRACTICAL IDEALISM IN U.S. FOREIGN POLICY 9 (2008).

[54] *See generally* RONALD E. POWASKI, AMERICAN PRESIDENTIAL STATECRAFT: FROM ISOLATIONISM TO INTERNATIONALISM (2017).

[55] ZAIDI, *supra* note 26.

In the eyes of the activists of their time, aviation and atomic energy were also potential vehicles for peace.[57] Such awesome power both demanded and made possible the creation of a new liberal world order. In the case of air power, civil aviation could bind the globe through trade and communication, while an international air force oversaw collective security;[58] after the devastation of the Second World War, international control of nuclear weapons could prevent catastrophic conflict, strengthening the fledgling United Nations.[59]

Both efforts failed. Yet their failures illuminate the politics of the time and the social currents that supported them, echoes of which lived on in the more modest oversight of civil aviation and atomic energy that did manifest, as well as in techno-globalist rhetoric that continues today.

In retrospect, those earlier efforts were part of a liberal internationalist arc that saw its beginnings in the nineteenth century.[60] In form, it saw the shift from bilateralism to multilateralism, notably including the first international organizations recognizable as such.[61] In substance, it encompassed efforts to restrain the use of force as well as to position international law as the conscience of the "civilized" world.[62] There is a darker aspect to this history, of course. Much as Martti Koskenniemi traced the throughline from colonialism to modern human rights,[63] proposals to use air power to maintain global order applied lessons learned in the governance of far-flung territories of empire.[64]

---

[56] *Id*. at 247.

[57] The language at times resonates with the techno-utopianism of today, with nuclear power plants described as the "cathedrals" of the twentieth century. *See, e.g.,* JACQUES LECLERCQ, THE NUCLEAR AGE (1986).

[58] ZAIDI, *supra* note 26, at 107-11.

[59] *Id*. at 204-09.

[60] *Id*. at 20-23. *See, e.g.,* MARK MAZOWER, GOVERNING THE WORLD: THE HISTORY OF AN IDEA (2012); GLENDA SLUGA, INTERNATIONALISM IN THE AGE OF NATIONALISM (2013).

[61] Simon Chesterman, David M. Malone, and Santiago Villalpando, *Introduction*, in THE OXFORD HANDBOOK OF UNITED NATIONS TREATIES 1 (Simon Chesterman, David M. Malone, and Santiago Villalpando eds., 2019).

[62] GERRIT W. GONG, THE STANDARD OF "CIVILIZATION" IN INTERNATIONAL SOCIETY (1984).

[63] MARTTI KOSKENNIEMI, THE GENTLE CIVILIZER OF NATIONS: THE RISE AND FALL OF INTERNATIONAL LAW 1870-1960 41. (2001)

[64] DAVID E. OMISSI, AIR POWER AND COLONIAL CONTROL: THE ROYAL AIR FORCE, 1919–1939 (1990).

The destructive power of these new technologies, along with enormous potential benefits — including, in the case of nuclear power, the prospect of electricity "too cheap to meter"[65] — saw the emergence of the technological internationalism of Zaidi's title. He tracks the emergence of a loose consensus that the new "machine age" required international governance through technical expertise.

Many early proposals for notionally global security forces were in fact premised on specific countries' armed services — notably one's own — operating under an international flag.[66] *The Economist* of the 1930s was wary of French ambitions in particular, but concluded that the case for internationalization was "overwhelming" as "[f]lying is supra-national in its very nature."[67] Discussion about the internationalization of aviation became more mainstream in the course of that decade, including when popularized in H.G. Wells' book and later film *The Shape of Things to Come*.[68]

The horrors of the Second World War led to the zenith of such aspirations, with plans for a post-war order including widespread discussion of an internationalized air force. Indeed, the 1944 meeting of the American Society of International Law featured detailed discussion of the merits of such a force, which could provide "an effective spearpoint for police action capable of immediate use by the international council at any point where aggression might occur or be threatened."[69] By the Dumbarton Oaks Conference some months later, however, enthusiasm for a truly international or exclusively aerial combat force had waned. The Soviet delegation raised the possibility of such an entity, but the compromise was that national contingents would be made available to the proposed Security Council "on its call and in accordance with a special agreement or agreements."[70] This was later formalized as Article 43 of the UN Charter, with a commitment that the specifics of the "numbers and types of forces, their degree of readiness and general location" would be negotiated "as soon as possible."[71] Despite several enforcement actions over the succeeding decades, not a single agreement as envisaged by the Charter has yet been concluded.[72]

---

[65] ROBERT L. BROWN, NUCLEAR AUTHORITY: THE IAEA AND THE ABSOLUTE WEAPON 55-61 (2015).

[66] ZAIDI, *supra* note 26, at 59-60.

[67] *Disarmament in the Air*, ECONOMIST, 25 February 1933.

[68] H.G. WELLS, THE SHAPE OF THINGS TO COME (1933); John S. Partington, *H.G. Wells and the World State: A Liberal Cosmopolitan in a Totalitarian Age*, 17(2) INTERNATIONAL RELATIONS 233 (2003).

[69] Quincy Wright, *Enforcement of International Law*, 38 PROC. AM. SOC'Y INT'L L. 77, 85 (1944).

[70] Washington Conversations on International Peace and Security Organization (Dumbarton Oaks Conference) (7 October 1944), *at* https://www.ibiblio.org/pha/policy/1944/441007a.html, Chapter VIII, Section B, para. 5.

The longevity of the technological internationalist view, at least among a certain class of intellectuals, can be seen in proposals that the latest technology offering tremendous benefits alongside real risks — AI — should also be governed in some measure by a cadre of global experts. The analogy with nuclear energy and the atomic bomb in particular is well-worn, having been embraced by academics,[73] leaders of technology companies,[74] and the Secretary-General of the United Nations himself.[75]

The limits of that analogy are obvious. Nuclear energy comprises a well-defined set of techniques using specific materials that are unevenly distributed around the world. AI, by contrast, is an amorphous term whose applications are extremely wide and difficult to contain. Atomic bombs are expensive to build and difficult to hide; weaponized AI promises to be neither.

Still larger problems may be political and structural. Politically, there is no appetite for anything remotely as elaborate as the International Atomic Energy Agency (IAEA) for AI.[76] Despite early enthusiasm on the part of the Secretary-General,[77] the Global Digital Compact, adopted by member states in late 2024, did not even mention an agency, calling instead for the UN Secretary-General to submit a proposal for an office that would draw upon existing resources to "facilitate system-wide coordination" in relation to AI.[78]

---

[71] Charter of the United Nations, done at San Francisco, 26 June 1945 (in force 24 October 1945), *at* http://https://www.un.org/en/about-us/un-charter/full-text, art. 43.

[72] The Military Staff Committee that was intended to "advise and assist" the Security Council on "the employment and command of forces placed at its disposal" remains a curiosity, continuing to meet despite having no formal agenda items since it reported in 1948 that it was unable to complete the mandate given to it two years earlier. *Id*., art. 47(1).

[73] Simon Chesterman, *Beyond Asimov's Three Laws: The Case for an International AI Agency*, ENGINEERING AND TECHNOLOGY, 4 August 2021; Gary Marcus and Anka Reuel, *The World Needs an International Agency for Artificial Intelligence, Say Two AI Experts*, ECONOMIST, 18 April 2023.

[74] Prarthana Prakash, *OpenAI's Sam Altman and Google's Sundar Pichai Are Now Begging Governments to Regulate the A.I. Forces They've Unleashed*, FORTUNE, 24 May 2023.

[75] Michelle Nichols, *UN Chief Backs Idea of Global AI Watchdog like Nuclear Agency*, BLOOMBERG, 13 June 2023.

[76] UN AI Advisory Body, Governing AI for Humanity: Final Report, *supra* note 14, para. 181.

[77] *Supra* note 75.

Structurally, international organizations like the UN are often ill-suited to — and often vehemently opposed to — the direct participation of private sector actors. In March 2024, for example, the General Assembly adopted its first ever resolution on regulating AI. The non-binding document calls on member states "and, where applicable, other stakeholders" not to use AI systems that pose undue risks to the enjoyment of human rights.[79] A few paragraphs later, the Assembly "*encourages* the private sector to adhere to applicable international and domestic laws."[80]

Zaidi concludes his book with the observation that technology through the ages has tapped into our existential hopes and fears, often emerging as "carriers of our dreams and nightmares."[81] With regard to AI, at least, waking up to address this latest challenge requires more than waiting for the United States to assert its role as a shining city on a hill, or hoping that the United Nations will somehow save humanity from its silicon creations.[82]

# 3   Reining in the Digital Robber Barons

Where Harding and Zaidi limit themselves to technological changes within living memory of many people, Acemoglu and Johnson take a more expansive view, aiming to encompass a millennium of progress in around half a thousand pages. For the paperback edition of their work, the two MIT professors — who were among the 2024 Nobel Prize winners *not* connected to or relying on AI[83] — summarize a key finding as being that "really bad outcomes are possible when deluded technology leaders are able to impose their messianic visions on society."[84]

---

[78] Global Digital Compact (United Nations, New York, September 2024), *at* https://www.un.org/techenvoy/global-digital-compact, para. 72.

[79] GA Res 78/265, UN Doc. A/78/265 (21 March 2024), para. 5.

[80] *Id*., para. 9.

[81] Zaidi, *supra* note 26, at 247.

[82] Former Secretary-General Dag Hammarskjöld is said to have quipped that organizations such as the UN were not created to take humanity to heaven, but to save it from hell.

[83] Jeanna Smialek, *Three Receive Nobel in Economics for Research on Global Inequality*, N.Y. Times, 14 October 2024.

None of the books discussed in this essay are entirely pessimistic about AI. Yet they are all deeply suspicious of the claims of techno-utopians that AI will yield nothing but benefits. The venture capitalist Marc Andreessen epitomized this Panglossian view in a document literally called "The Techno-Optimist Manifesto," asserting that the productivity boost from technologies such as AI "drives wages *up*, not down. This is perhaps the most counterintuitive idea in all of economics, but it's true, and we have 300 years of history that proves it."[85] (It's not, and he doesn't.) Amazon's Jeff Bezos pushed this into caricature in his final letter to shareholders as Chief Executive Officer. Trying to head off criticisms of worker treatment that had led to fights over unionization and criticism of worker safety, Bezos promised a new commitment to "a better vision for our employees' success." The form this vision would take could have been lifted from *Brave New World*, including "new automated staffing schedules that use sophisticated algorithms to rotate employees among jobs that use different muscle-tendon groups to decrease repetitive motion."[86]

It is true that most of us are materially better off than our ancestors. The reason is only partly technology, however. Acemoglu and Johnson argue that it is also because "citizens and workers in early industrial societies organized, challenged elite-dominated choices about technology and work conditions, and forced ways of sharing the gains from technical improvements more equitably."[87]

They interrogate the idea that technological progress is always economically progressive. On the contrary, they argue, the last thousand years saw serial fights over the direction of technology and the type of progress, with accompanying winners and losers.[88] Their main target is the "productivity bandwagon" — the assumption that new machines that increase productivity will also increase wages and benefit everyone, not merely the entrepreneurs and owners of capital.[89] The book considers the rise of modern agriculture, the industrial revolution, and the emergence of digital technologies. Each of these periods saw a tension between who led change and who benefited from it.

---

[84] ACEMOGLU AND JOHNSON, *supra* note 27, at xiii.

[85] Marc Andreessen, The Techno-Optimist Manifesto (Andreessen Horowitz, Menlo Park, CA, 16 October 2023), *at* https://a16z.com/the-techno-optimist-manifesto/.

[86] Jeff Bezos, 2020 Letter to Shareholders (Amazon, Seattle, WA, 16 April 2021), *at* https://www.aboutamazon.sg/news/company-news/2020-letter-to-shareholders. Bezos continues to serve as Executive Chairman of Amazon.

[87] ACEMOGLU AND JOHNSON, *supra* note 27, at 7.

[88] *Id*. at 34.

The main battleground is domestic, including traditional political institutions and organized labor.[90] A signal inflection point has been the move away from technology that created new tasks and opportunities to the automation of work and cutting labor costs, which they attribute to "lack of input and pressure from workers, labor organizations, and government regulation."[91] In some ways, it is ironic that digital technology played such a role in this context. Echoing Harding, they note that the early days of the computer revolution were defined by decentralization and freedom, bordering on anarchy.[92] An observer might have predicted that subsequent decades would "further bolster countervailing powers against big business, create new productive tools for workers, and lay the foundations of even stronger shared prosperity."[93]

Instead, digital technologies became "the graveyard of shared prosperity."[94] Though other factors were at play — notably globalization and the weakening of the labor movement in the United States in particular — the change in the direction of technology from the 1970s onward was decisive. Much of the energy and creativity has gone into replacing humans through automation and surveilling those workers who remain.[95]

In opposition to machine intelligence, Acemoglu and Johnson encourage the pursuit of machine *usefulness*, meaning the search for ways in which AI could better complement human workers, rather than merely replacing them.[96] Their call to action can be vague at times: "altering the narrative, building countervailing powers, and developing technical, regulatory, and policy solutions to tackle specific aspects of technology's social bias."[97] The final chapter does offer a raft of policy reforms, but their central message is that passively accepting the social costs of technology is not the only option. The Gilded Age of the late nineteenth century also saw periods of rapid technological change — railways and oil, steel and finance — along with rent-seeking from the robber barons of the era. Checking the massive inequality that followed depended on action by civil society, notably journalists (the original muckrakers) and organized labor, driving a political movement that in turn led to greater regulation and efforts to rein in the power of corporate titans through antitrust and campaign finance reform.[98]

---

[89] *Id*. at 14.

[90] Though somewhat interested in other countries' approaches, the global nature of the problem is largely ignored, except to the extent that "international cooperation" might be necessary to implement taxation measures proposed. *Id*. at 415. The authors acknowledge the "central question" of whether redirecting technology in the West "would be of any use if China continues to pursue automation and surveillance." They conclude that it the answer is "likely yes," on the basis that China is a "follower" in most frontier technologies. *Id*. at 395.

[91] *Id*. at 37.

[92] *Id*. at 253-54.

[93] *Id*. at 255.

[94] *Id*..

[95] *Id*. at 297-338.

Unfortunately, such institutions barely exist today. Though there is impressive reporting on the power of the modern digital robber barons, the decline of journalism as an industry has been exacerbated by the rise of disinformation supercharged by generative AI.[99] "Democracy dies in darkness," Acemoglu and Johnson note, alluding to the mission statement adopted by the *Washington Post* at the start of President Donald Trump's first term in office.[100] "But it also struggles under the light provided by modern artificial intelligence."[101] It is a telling indicator of the impact of market forces that the *Post* more recently abandoned that aspirational slogan for the less confrontational and market-friendly "Riveting Storytelling for All of America."[102]

---

[96] *Id*. at 327-32.

[97] *Id*. at 38.

[98] *Id*. at 383-86.

[99] Kokil Jaidka et al., *Misinformation, Disinformation, and Generative AI: Implications for Perception and Policy*, forthcoming Digit. Gov't: Rsch. & Prac. (2024).

[100] Paul Farhi, *The Washington Post's New Slogan Turns Out to Be an Old Saying*, Wash. Post, 24 February 2017. The phrase about government accountability was popularized by journalist Bob Woodward, who credited it in turn to a first amendment case. *See Detroit Free Press v. Ashcroft*, 303 F.3d 681 (6th Cir., 2002) ("Democracies die behind closed doors.").

[101] Acemoglu and Johnson, *supra* note 27, at 352.

[102] Benjamin Mullin, *The Washington Post's New Mission: Reach "All of America"*, N.Y. Times, 16 January 2025.

With respect to organized labor, the marginalization of the working class more generally has been linked to the rise of authoritarian politics in the United States and elsewhere.[103] A rare example of unions mobilizing workers to limit the impact of AI came from the very elites typically derided by populist politicians: Hollywood. The Writers Guild of America strike in 2023 sought, among other things, to limit the use of AI to helping with research or facilitating scriptwriting — rather than replacing writers.[104] The strike gained extraordinary publicity in part because it naturally impacted the media landscape, but also because workers effectively exercising power in the United States is very much the exception rather than the norm.

Acemoglu and Johnson's other prescriptions include measures to alter the incentives for technology companies. Taxes currently encourage automation, for example. In the United States, a company investing in automation equipment or software pays a fifth of the tax it would face if hiring workers to perform the same tasks.[105] They do not support an automation tax, but suggest limiting such incentives, along with denying patent protection for surveillance technologies.[106]

A larger problem of incentives is the extent to which revenue streams for tech companies currently rely on advertising as opposed to, say, subscriptions. Subscription models encourage companies to curate quality experiences for their users; advertising generally rewards the *quantity* of engagement. Few users are willing to pay for content, however, giving rise to the model of funding that content by monetizing the personal data of those users. That model was summed up in the pithy phrase that "if something is free, you are the product,"[107] later termed "surveillance capitalism" by Shoshana Zuboff.[108]

---

[103] *See generally* MARC J. HETHERINGTON AND JONATHAN DANIEL WEILER, AUTHORITARIANISM AND POLARIZATION IN AMERICAN POLITICS (2009); THE GLOBAL RISE OF AUTHORITARIANISM IN THE 21ST CENTURY: CRISIS OF NEOLIBERAL GLOBALIZATION AND THE NATIONALIST RESPONSE (Berch Berberoglu ed., 2021).

[104] Alexandra Curren, *Digital Replicas: Harm Caused by Actors' Digital Twins and Hope Provided by the Right of Publicity*, 102(1) TEX. L. REV. 155 (2023).

[105] ACEMOGLU AND JOHNSON, *supra* note 27, at 406.

[106] *Id*. at 403.

[107] This idea is sometimes traced back to a 1973 video by the artist Richard Serra and Carlota Fay Schoolman. Richard Serra and Carlota Schoolman, Television Delivers People (Museum of Modern Art, New York, 1973), *at* https://www.moma.org/collection/works/118185 ("You are the end product of t.v."). *Cf.* Claire Wolfe, *Little Brother Is Watching You: The Menace of Corporate America*, LOOMPANICS 1999 ("you're not the customer any more. You're simply a "resource" to be managed for profit."). *See also* Kashmir Hill, *You Are the Product*, N.Y. TIMES, 30 May 2021.

As a means of shifting these incentives, Acemoglu and Johnson propose a "nontrivial digital advertising tax."[109] Though unable to put a price on this, they are optimistic that it would encourage "alternative business models." In addition to subscriptions, which fund companies such as Netflix and the *New York Times*, those alternative models include sites like Wikipedia, which draws on the wisdom of the crowd and structured governance institutions to develop and maintain standards. More importantly, it is funded through a not-for-profit foundation.[110]

The limits of that approach were on display in the corporate drama that unfolded in OpenAI, the company behind ChatGPT. It was established as a non-profit in 2015, heralded with lofty statements as to how this status enabled it to "benefit humanity as a whole, unconstrained by a need to generate financial return."[111] As the costs of training the large language model that became its signature product escalated, the company announced that it would adopt a "capped-profit" structure, allowing it "to rapidly increase our investments in compute and talent."[112] The contradiction between these two worldviews unfolded in the spectacle of the not-for-profit Board firing CEO Sam Altman in November 2023 — only for him to be reinstated days later and the Board itself being replaced. As various commentators pointed out: "The money always wins."[113]

---

[108] Shoshana Zuboff, The Age of Surveillance Capitalism: The Fight for a Human Future at the New Frontier of Power (2019). *See also* Nick Srnicek, Platform Capitalism (2016); Tarleton Gillespie, Custodians of the Internet: Platforms, Content Moderation, and the Hidden Decisions that Shape Social Media (2018); Yanis Varoufakis, Technofeudalism: What Killed Capitalism (2023).

[109] Acemoglu and Johnson, *supra* note 27, at 413-14.

[110] *Id*. at 379. Perhaps tellingly, for an American audience, Acemoglu and Johnson do not bother to mention a third obvious possibility: public funding.

[111] Greg Brockman and Ilya Sutskever, Introducing OpenAI (OpenAI, 11 December 2015), *at* openai.com/blog/introducing-openai.

[112] Greg Brockman and Ilya Sutskever, OpenAI LP (OpenAI, 11 March 2019), *at* openai.com/blog/openai-lp.

[113] Charlie Warzel, *The Money Always Wins*, Atlantic, 21 November 2023. That conclusion was underlined by the subsequent departure from OpenAI of Jan Leike and Ilya Sutskever in May 2024, hollowing out the "super-alignment" team at the most influential AI company on the planet. Will Knight, *OpenAI's Long-Term AI Risk Team Has Disbanded*, Wired, 17 May 2024. To the extent that the OpenAI saga reflected debates over the existential threat that might be posed by AI, it was a victory of the capitalists over the catastrophists. Kevin Roose, *A.I. Belongs to the Capitalists Now*, N.Y. Times 2023.

# 4   The Tech-Industrial Complex

So, what is to be done? If companies cannot be trusted to self-regulate, if governments are unwilling to legislate, and if international organizations are unable to do more than coordinate — who or what might help mitigate the risks and more evenly distribute the benefits of AI?

Returning to Harding's titular exhortation, the first answer is, of course, us. Users can choose not to support companies that ignore safety or exacerbate inequality. The problem is that individual users have trivially little leverage over companies whose business model is premised in part on hiding that lack of agency from consumers.[114] The tragedy of AI governance lies in that inverse relationship between leverage and interest: users have interest but no leverage; tech companies have leverage but no interest in constraining their own behavior if it means thereby limiting their profits.[115]

Just as organized labor offered glimmers of hope in increasing workers' bargaining power, organized users might have a greater say in how technology is developed and deployed. Global privacy movements, for example, shifted markets at least modestly, as reflected in the rise of privacy-by-design and personal data protection being seen by some companies as a market differentiator.[116] It is conceivable that similar norms might emerge in the AI space, perhaps along the lines of "responsible" AI that is more trustworthy and less prone to hallucinations, or more "open" in the sense of greater transparency as to how decisions are made and how models are trained.[117]

---

[114] *See, e.g.,* CLAUDIO CELIS BUENO, THE ATTENTION ECONOMY: LABOUR, TIME, AND POWER IN COGNITIVE CAPITALISM (2017); JENNY ODELL, HOW TO DO NOTHING: RESISTING THE ATTENTION ECONOMY (2019); TIMOTHY AYLSWORTH AND CLINTON CASTRO, KANTIAN ETHICS AND THE ATTENTION ECONOMY: DUTY AND DISTRACTION (2024); KAREN NELSON-FIELD, THE ATTENTION ECONOMY: A CATEGORY BLUEPRINT (2024).

[115] *See* Simon Chesterman, *The Tragedy of AI Governance*, *in* CONTEMPORARY DEBATES IN THE ETHICS OF ARTIFICIAL INTELLIGENCE forthcoming (Atoosa Kasirzadeh, Sven Nyholm, and John Zerilli eds., 2025).

[116] GARRETT JOHNSON, ECONOMIC RESEARCH ON PRIVACY REGULATION: LESSONS FROM THE GDPR AND BEYOND (2024).

[117] Tita Alissa Bach et al., *Insights into Suggested Responsible AI (RAI) Practices in Real-World Settings: A Systematic Literature Review*, forthcoming AI AND ETHICS (2025).

Another form of transparency would be as to the costs of AI, notably its environmental impact.[118] Various tech companies — and some countries — have announced that their investments in AI mean that they are giving up on climate targets, though they have largely refrained from passing on these costs to consumers.[119] More information about the costs of AI, either through moves to subscription models, or at least revealing the electricity and water consumed when using the latest AI systems, might influence user and therefore corporate behavior.

Market mechanisms will not be enough, however. Days before he left office, President Joe Biden spoke of the emergence of a "tech-industrial complex,"[120] echoing Dwight Eisenhower's own valedictory address on the "military-industrial complex."[121] While Eisenhower was concerned about the influence of the arms industry on military procurement and defense policy,[122] Biden warned that "an oligarchy is taking shape in America of extreme wealth, power and influence that really threatens our entire democracy."[123]

In the wake of the global financial crisis of 2007-2008, one of the lessons learned was that if certain banks were "too big to fail," then it meant that they were too big in the first place.[124] Echoing earlier battles from the Gilded Age, there is a strong argument that tech companies — or tech entrepreneurs — that are too big to regulate are too big, period. There have, of course, been efforts to break up those companies. The U.S. Justice Department is currently suing Google[125] and Apple,[126] while the Federal Trade Commission has ongoing actions against Amazon,[127] having unsuccessfully brought actions against Microsoft[128] and Meta.[129] In addition to its own antitrust actions,[130] the European Union has linked size with more elaborate obligations and reporting requirements for "gatekeepers" under the Digital Markets Act[131] and "very large" online platforms and search engines under the Digital Services Act.[132] Only China, however, has successfully broken up tech companies in a purge lasting from 2020 to 2023, wiping trillions of dollars wiped off the share value of those companies,[133] with Alibaba broken into six new entities.[134] These were costs that Beijing was willing to bear, but at which Washington or Brussels might balk, particularly given President Trump's new chumminess with the tech elite.

---

[118] *See* CRAWFORD, *supra* note 6; Measuring the Environmental Impacts of Artificial Intelligence Compute and Applications: The AI Footprint (OECD Digital Economy Papers, Paris, 15 November 2022); Nir Kshetri, *The Environmental Impact of Artificial Intelligence*, 26(3) IT PROFESSIONAL 9 (2024).

[119] *See generally* Stephanie Jamison et al., Destination Net Zero: Fast-Tracking Progress (Accenture, Dublin, 2024), *at* https://www.accenture.com/us-en/insights/sustainability/destination-net-zero.

[120] Erica L. Green, *In Farewell Address, Biden Warns of an 'Oligarchy' Taking Shape in America*, N.Y. TIMES, 15 January 2025.

[121] President Dwight D. Eisenhower's Farewell Address (17 January 1961), *at* https://www.archives.gov/milestone-documents/president-dwight-d-eisenhowers-farewell-address.

[122] JAMES LEDBETTER, UNWARRANTED INFLUENCE: DWIGHT D. EISENHOWER AND THE MILITARY-INDUSTRIAL COMPLEX (2011). *Cf.* GORDON ADAMS, THE POLITICS OF DEFENSE CONTRACTING: THE IRON TRIANGLE (1981).

[123] Green, *supra* note 120.

[124] *Greenspan Calls to Break Up Banks "Too Big to Fail"*, N.Y. TIMES, 15 October 2009.

[125] *United States v. Google LLC*, 1:23-cv-00108 (E.D. Va., 2023).

As for international institutions, another Eisenhower speech from the very first year of his presidency suggests the possibilities and limitations. By 1953, the technological internationalist moment had passed and the prospect of international control of nuclear weapons faded. Eisenhower proposed an alternative in his "Atoms for Peace" address to the UN. If the earlier idea had been utopian, this was idealistic in a different way: instead of concentrating nuclear materials and expertise in a supranational body, they would be disseminated across the globe — encouraging states to use them for peaceful purposes, in exchange for commitments to renounce the search for the bomb.[135]

---

[126] *U.S and Plaintiff States v. Apple Inc.* (available at <http://https://www.justice.gov/atr/case/us-and-plaintiff-states-v-apple-inc>).

[127] Amazon.com, Inc. (Amazon eCommerce) (Federal Trade Commission, Washington, DC, 2024), *at* https://www.ftc.gov/legal-library/browse/cases-proceedings/1910129-1910130-amazoncom-inc-amazon-ecommerce.

[128] Kellen Browning, David McCabe, and Karen Weise, *Judge Rejects F.T.C. Delay of $70 Billion Microsoft-Activision Deal*, N.Y. TIMES, 11 July 2023.

[129] Diane Bartz, *FTC Withdraws from Adjudication in Fight with Meta over Within Deal*, REUTERS, 11 February 2023.

[130] *See* ANTONIO MANGANELLI AND ANTONIO NICITA, REGULATING DIGITAL MARKETS: THE EUROPEAN APPROACH (2022); CARLO MARIA COLOMBO, KATHRYN WRIGHT, AND MARIOLINA ELIANTONIO, THE EVOLVING GOVERNANCE OF EU COMPETITION LAW IN A TIME OF DISRUPTIONS: A CONSTITUTIONAL PERSPECTIVE (2024).

[131] Regulation (EU) 2022/1925 of the European Parliament and of the Council of 14 September 2022 on Contestable and Fair Markets in the Digital Sector and Amending Directives (EU) 2019/1937 and (EU) 2020/1828 (Digital Markets Act) 2022 (EU). Six companies have been so designated: Alphabet (the parent company of Google), Amazon, Apple, ByteDance (which owns TikTok), Meta, and Microsoft.

[132] Regulation (EU) 2022/2065 of the European Parliament and of the Council of 19 October 2022 on a Single Market for Digital Services and amending Directive 2000/31/EC (Digital Services Act) 2022 (EU).

[133] Lilian Zhang, *A Timeline of China's 32-month Big Tech Crackdown that Killed the World's Largest IPO and Wiped Out Trillions in Value*, SOUTH CHINA MORNING POST, 15 July 2023.

[134] *Alibaba Starts a Spate of Spinoffs*, INDIA BUSINESS JOURNAL, June 2023.

The analytical, political, and structural limits of this analogy have been touched on earlier.[136] The biggest difference between attempts to control nuclear power in the 1950s and AI today, however, may be the historical context. For even as Eisenhower spoke in New York, the effects of the nuclear blasts on Hiroshima and Nagasaki were still being felt.[137] The "dread secret" of those weapons, he warned, was no longer confined to the United States and no longer containable. To do nothing was to accept the hopeless finality that "two atomic colossi are doomed malevolently to eye each other indefinitely across a trembling world."[138]

There is, at present, no such threat from AI — nor is there comparably visceral evidence of its potential for harm. It is possible that concerns are overblown. Or, as some would argue, AI itself may help solve these and sundry other problems.[139] If not, however — if the unchecked power of tech companies and their silicon sovereigns is unable to be constrained or contained by users, the market, or states in thrall to the new mammon — then global institutions that might have helped to prevent the first true AI emergency may be too late to avert the second.

---

[135] Address by Mr. Dwight D. Eisenhower, President of the United States of America, to the 470th Plenary Meeting of the United Nations General Assembly (Atoms for Peace) (United Nations, New York, 8 December 1953), *at* https://www.iaea.org/about/history/atoms-for-peace-speech.

[136] *See supra* notes 73-78.

[137] Lesley M.M. Blume, Fallout: The Hiroshima Cover-Up and the Reporter Who Revealed It to the World (2020).

[138] Atoms for Peace, *supra* note 135.

[139] *See, e.g.,* Lawrence Lessig, Code: Version 2.0 ([1999] 2006); Karen Yeung, *"Hypernudge": Big Data as a Mode of Regulation by Design*, 20 Information, Communication & Society 118 (2017); Marco Almada, *Regulation by Design and the Governance of Technological Futures*, 14(4) Eur. J. Risk Reg. 697 (2023).